\documentclass[11pt]{article}
\usepackage{amsmath}  
\usepackage{amsfonts}
\usepackage{geometry}              
\usepackage{graphicx}
\usepackage{amssymb}
\usepackage{epstopdf}
\newcommand{\dd}{\delta_\nu}

\newcommand{\be}{\begin{equation}}
\newcommand{\ee}{\end{equation}}
\newcommand{\beq}{\begin{eqnarray}}
\newcommand{\eeq}{\end{eqnarray}}
\newcommand{\beqs}{\begin{eqnarray*}}
\newcommand{\eeqs}{\end{eqnarray*}}

\title{Casimir energy for a spherical boundary  within the surface impedance approach:
 Obtaining negative values}

\author{Luigi Rosa$^{1,2}$} 

\date{}                                           

\begin{document}
\maketitle

\begin{center}
{\em $^1$ Dipartimento di Scienze Fisiche, Universit\`a Federico II,\\
Complesso Universitario di Monte S. Angelo, Via Cintia Edificio 6, 80126 Napoli, Italy\\
$^2$INFN, Sezione di Napoli, Complesso Universitario di Monte S. Angelo, Via Cintia Edificio 6, 80126 Napoli, Italy}
\end{center}

\begin{abstract}
We compute the Casimir Energy of a spherical region using a { Surface Impedance } approach. We characterize the Surface Impedance  of the boundary using plasma model. 
Exact analytical formulae are obtained by means of  the  zeta function regularization method  and all the divergencies are explicitly computed. We find that it is possible to have negative Casimir energy for some range of values of the relevant parameter $y_a=\omega_p a \sqrt{\epsilon\mu}$. Limits of applicability of the model are discussed.

\vskip1truecm
{\bf Pacs: 12.20.Ds, 12.39.Ba, 03.70.+k}
\end{abstract}

\section{Introduction}

The interest in the Casimir energy in spherical bodies is enormous. The first attempt of Casimir to furnish a model for the electron \cite{cas53} in which an {\em attractive} Casimir force would balance the electrostatic self-repulsion showed to be unsuccesful after the finding of Boyer  \cite{boyer68} that the Casimir force in a sphere is repulsive. The interest in this configuration is still very strong and the use of advanced computational tools has simplified very much the calculation (see \cite{bord09} and references therein). Here we make another small step, aimed at further simplifying  the procedure, implementing the Surface Impedance approach in this kind of problems.

Surface Impedance (SI)  can result very useful in modeling the behavior of a non ideal surface. It is defined through the equation \cite{strat41,landau8,jack98,moste84}:
\be
\mathbf{E}_{tan}|_a={\cal{Z}}\left(\hat{n}\times\mathbf{B}\right)|_a \label{eq:defz}
\ee
with $\vec{n}$ the outward normal to the surface. This formula relates the tangential fields outside the material surface and all the characteristics of the material are taken into account through the values of $\cal{Z}$. Eq. (\ref{eq:defz}) can be seen as the defining equation for $\cal{Z}$ and it can be applied to arbitrary materials even when a description in terms of dielectric permittivity cannot be given \cite{esqui03,geklim03}. In particular it can result very useful in describing the transition from attractive to repulsive behavior \cite{kenn02,rosa10} when computing Casimir energy \cite{cas48,bomu01} and, in this respect, this paper can be seen as the natural prosecution of \cite{rosa10}. (As far as we know the first time SI has been used in connection with Casimir energy was in \cite{moste84} ).

The approach is very general and powerful and, in our opinion, it can treat, in a relatively easy manner, problems with spherical boundaries: a dielectric sphere \cite{bart99}, the bag model of quantum chromodynamics \cite{weiss74,chod74}, quantum gravity and cosmology \cite{gesp94,gesp99}, boundaries induced by topological parameters \cite{teta} etc. 
Obviously the possibility of using an equation like (\ref{eq:defz}) must be treated carefully case by case, here we make a first attempt having in mind {\em  "standard cases"} like electromagnetic field in a sphere and the MIT bag model of QCD \cite{chod74,plunien}. It is important to stress from the very beginning that within this model the renormalization procedure is not completely settled and the model is still physically unsatisfactory \cite{milton01,bord09}.
To regularize the Casimir energy we will use the  zeta function  regularization
\cite{lese96,eliza95,bord96,bord961,kirsten02,bord09}. 

The paper is organized as follows: in Sec. II the  zeta function  regularization approach is shortly revised for the convenience of the readers. In Sec. III the relevant formulae are derived and the Casimir energy is computed, finally in Sec. IV we analyse the results. 

\section{Casimir Energy and Zeta function regularization }
The Casimir energy is the vacuum expectation value of the Hamiltonian operator on the ground state:\be
E_{Cas}=< 0| H |0 >=\frac{1}{2}\sum_J\left(E_J\right)
\ee
where $E_J$ are the energy eigenvalues labelled by some general index $J$. The sum is, in general, divergent and regularization is necessary. In the $\zeta$  function regularization scheme one defines a new quantity $\zeta_{H}(s)$ as
\be
E_{Cas}=\lim_{s\to-1/2}\frac{\mu^{2s+1}}{2}\sum_J \left(E_J^2\right)^{-s}=:\lim_{s\to-1/2}\mu^{2s+1}\zeta_H(s)
\ee
where $\zeta_H(s)$ is the $\zeta$ function relative to the differential operator connected to the operator $H$. The parameter $\mu$ is an arbitrary parameter introduced for dimensional reasons, it will disappear on removing the regularization in the limit $s\to-1/2$. In general no explicit expression for the eigenvalues $E_J$ exists, however one can use the argument principle to represent the sum over the eigenvalues as a contour integral in the complex plane \cite{vanka68,arfken}:
\be
\sum_k g(a_k)-\sum_mg(b_m)=\frac{1}{2\pi i}\oint_\gamma g(z) d \log(\Delta(z))
\ee
where $\gamma$ is a closed contour containing all the zeros $a_n$ and poles $b_n$ of the function $\Delta(z)$ assumed analytic in $\gamma$ and $g(z)$ is some analytic function inside $\gamma$.  Thus if $\Delta(z)$ is such that $\Delta(\omega_n)=0$, $\omega_n$ being the eigenvalues of our problem, and has no pole, the sum over $\omega_n$ can be obtained as a contour integral. In this case $\Delta(z)$ is called, for obvious reasons, the mode-generating function. In conclusion $\zeta_H$ can be written as a contour integral of some mode-generating function \cite{gesp97}:
\be
\zeta_H=\sum_J \frac{1}{2\pi i}\oint_\gamma g(z) d \log(\Delta_J(z)) \label{eq:zh}
\ee
where the sum over $J$ takes into account possible degeneracy of the eigenvalues.

\section{The Casimir Energy for a Sphere}
In the following we will concentrate on the case of an electromagnetic field in a  sphere of radius $a$. We will characterize the boundary by means of its  Surface Impedence.
Assuming a time dependence $e^{-\i\omega t}$  the electric and magnetic fields in the interior of the sphere can be written in the form \cite{jack98,bord09,mie,bowcas}

\begin{eqnarray*}
\mathbf{E}   &=&\sum_{l=1}^{\infty }\frac{i}{rk  }a^{TE}    \left[ i\hat{\mathbf{n}}j_{\nu }(k  r)Y_{lm}(\theta ,\phi )+(k  rj_{\nu}(k  r))^{\prime }\hat{\mathbf{n}}\times \mathbf{X}_{lm}\right] +a^{TM}    j_{\nu }(k  r)\mathbf{X}_{lm} \\
\mathbf{H}   &=&\sum_{l=1}^{\infty }\frac{k  }{\omega \mu   }\left\{a^{TE}    j_{\nu }(k  r)\mathbf{X}_{lm}-i\frac{a^{TM}    }{rk  }\left[ i \hat{\mathbf{n}}j_{\nu }(k  r)Y_{lm}(\theta ,\phi )+(k  rj_{\nu}(k  r))^{\prime }\hat{\mathbf{n}}\times \mathbf{X}_{lm}\right] \right\},
\end{eqnarray*}
where $Y_{lm}$ and $\mathbf{X}_{lm}$ are the scalar and the vectorial spherical harmonics respectively, 
$k=\sqrt{\epsilon\mu}\omega$,  $j_{\nu }(x)=\sqrt{\frac{\pi }{2x }}J_{l+1/2}(x)$ are the spherical Bessel functions \cite{jack98,abste}, and $(xf(x))^{\prime }\equiv \frac{d}{dx}(xf(x))$. 

Imposing boundary conditions: Eq. (\ref{eq:defz}), we find the equations for the $TE$ and $TM$ modes:
\beq
\Delta^{TE}_\nu  (x) &:=&\left[\frac{i}{ka}\left(ka j_{\nu }(ka)\right)'-{\cal{Z}} j_{\nu }(ka)\right] a^{TE}   = 0 \\
\Delta^{TM}_\nu  (x) &:=&\left[-{\cal{Z}}\frac{i}{ka}\left(ka j_{\nu }(ka)\right)'+j_{\nu }(ka) \right]a^{TM}   = 0,
\eeq 
where $\Delta^{TE}_\nu  $ and $\Delta^{TM}_\nu  $ are our mode-generating functions.
In this case $\zeta_H$ can be given in the following form \cite{kirsten02,bord09}:
\be
\zeta_H(s)=\sum_{n=0}^\infty\sum_{l=0}^{\infty}\left( l+\frac{1}{2}\right)(\omega^2_{l,n}+m^2)^{-s}
\ee
where $\omega_{l,n}$ are the eigenmodes, $\hbar=1,c=1$ is assumed. We introduced a (fictitious) mass parameter that we will let go to zero in the end of the calculation so to avoid some problems with the integral representation of the $\zeta$ function. Using formula (\ref{eq:zh}), with a suitable choice of the contour $\gamma$, $\zeta_H$ can be written \cite{kirsten02}
\be
\zeta_H(s)=\sum_{l=1}^{\infty}\nu \int_\gamma \frac{dk}{2\pi i} \left( k^2+m^2\right)^{-s}
\frac{\partial}{\partial k} \log{\left[ \Delta^{TE}_\nu  (k a)+\Delta^{TM}_\nu  (k a)\right]}.
\ee
with $\nu=l+1/2$. Shifting the integration contour along the imaginary axis and using  $j_l(x)=\sqrt{\frac{\pi}{2x}}J_\nu(x)$ and $J_\nu(i x)=e^{i\pi\nu} J_\nu(-i x)$ and $J_\nu(i x)=e^{i\nu\frac{\pi}{2} }I_\nu(x)$, $I_\nu(x)$ being the modified Bessel functions \cite{abste},  we obtain the following expression valid in the strip $1/2<\Re(s)<1$.
\beq
\zeta_H(s)  &=&\frac{\sin(\pi s)}{\pi}\sum_{\nu=3/2}^\infty \nu \int_m^\infty dk\left[ k^2-m^2\right]^{-s}\frac{\partial}{\partial k}\left[\log\left(k^{-2\nu}\tilde{\Delta}^{TE}_\nu  (k a)\tilde{\Delta}^{TM}_\nu  (k a)\right)\right] \label{eq:zh2} \\
&=& \frac{\sin(\pi s)}{\pi}\sum_{\nu=3/2}^\infty \nu \int_{m a/\nu}^\infty dy\left[ 
\left(\frac{y\nu}{a}\right)^2-m^2\right]^{-s}\frac{\partial}{\partial y}\left[\log\left(y^{-\nu}\tilde{\Delta}^{TE}  ( y\nu)\right)+\log\left(y^{-\nu}\tilde{\Delta}^{TM}  (y\nu)\right)\right] \nonumber \\
&=:&\zeta^{TE}  (s)+\zeta^{TM}  (s)
\eeq
where,
after rotation to the imaginary axes, the mode-generating functions are given by: 
\beq
\tilde{\Delta}^{TE}_\nu  (x) &=&I_\nu(x)\left[ 1-\frac{{\cal{Z}}({i x/a})}{2x} \right]-{\cal{Z}}({i x/a})\dot{I}_\nu(x) \\
\tilde{\Delta}^{TM}_\nu  (x) &=&I_\nu(x)\left[ \frac{1}{2}-{\cal{Z}}({i x/a})x \right]+x\dot{I}_\nu(x) 
\eeq
It is clear, now, the role played by the mass $m$. Indeed representation (\ref{eq:zh2}) is defined for no values of $s$ if $m=0$. The procedure can be modified for $m=0$ but it is more difficult \cite{eliza93}.
Unfortunately we need the zeta function computed to the left of the strip $1/2<\Re(s)<1$. Now, the restriction $1/2<\Re(s)$ is due to the behavior of integrand as $k\to\infty$. The general technique \cite{kirsten02} to overcome this problem is to add and subtract the asymptotic term 
$$\left[\log\left(y^{-\nu}\tilde{\Delta}^{TE}  ( y\nu)\right)+\log\left(y^{-\nu}\tilde{\Delta}^{TM}  (y\nu)\right)\right]_{\nu\to\infty}$$
to the integrand so to move the strip of convergence to the left. If we are able to compute analytically the asymptotic term alone and at the same time to treat the remaining term at least numerically ({\em  but, in general, it can be made very small by considering higher and higher terms in the asymptotic expansion}) we can obtain the required analytical continuation.

To be more specific, let us define $\zeta_N(s)$ by the following equality
\be
\left(\zeta_H(s) -\zeta_H^{asym}(s) \right)+ \zeta_H^{asym}(s) =:\zeta_{N}(s)+\zeta_H^{asym}(s).
\ee
In the following we will give a representation of $\zeta_H^{asym}(s)$ in terms of known functions and valid in the region of interest of the complex plane. Obviously we have to choose an expression for $\cal{Z}$ to characterize the properties of the medium. We will use  the plasma model; thus 
$${\cal{Z}}\left(i \frac{y\nu}{a} \right)=\frac{y}{\sqrt{\delta_\nu^2+y^2}}=:\frac{y}{x_{\dd}}$$
with  $x_{\dd}=\sqrt{\delta_\nu^2+y^2}$, $\delta_\nu=\frac{y_a}{\nu}$ and $y_a=a \sqrt{\epsilon\mu}\omega_p$,  $\omega_p$ being the plasma frequency of the material constituting the surface.
At this point the procedure is quite standard \cite{ lese96,eliza95,kirsten02,bord09}
we need the asymptotic values of $\tilde{\Delta}^{TE}  $ and $\tilde{\Delta}^{TM}  $ for $\nu\rightarrow \infty$ with $k/\nu$ fixed. To this extent we use the uniform asymptotic expansions of ${I}_\nu(x)$ and $\dot{I}_\nu(x)$ \cite{olver,abste}:
\beq
I_\nu(\nu x) &=&\frac{1}{\sqrt{2\pi \nu}}\frac{e^{\nu \eta}}{\left(1+y^2\right)^{1/4}}\left[1+\sum_{k=1}^\infty\frac{u_k(t)}{\nu^k}\right]  \\
\dot{I}_\nu(\nu x) &=&\frac{1}{\sqrt{2\pi \nu}}{e^{\nu \eta}}\frac{\left(1+y^2\right)^{1/4}}{y}\left[1+\sum_{k=1}^\infty\frac{v_k(t)}{\nu^k}\right]  
\eeq
with
\beqs
t &=& \frac{1}{\sqrt{1+y^2}}, \text{  and  }  \eta=\sqrt{1+y^2}+
\ln\left(  \frac{y}{1+\sqrt{1+y^2}}\right); \\
u_0(t) &=&1,\text{  and  }  u_{k+1}(t)=\frac{t^2(1-t^2)}{2}u_k'(t)+\frac{1}{8}\int_0^t dz(1-5 z^2)u_k(z);   k=0,1,2\ldots \\
v_0(t) &=&1,\text{  and  } v_k(t)=u_k(t)-t(1-t^2)\left[ \frac{1}{2}u_{k-1}(t)+t u'_{k-1}(t)\right];   k=0,1,2\ldots  
\eeqs

\subsection{TE-Modes}
Inserting the asymptotic expansions of $I_\nu$ and $\dot{I}_\nu$ in (\ref{eq:zh2})  we obtain:
\beq
\zeta_{TE}^{asym}(s)
&=&\frac{\sin(\pi s)}{\pi}\sum_{\nu=3/2}^\infty \nu \int_{m a/\nu}^\infty dy\left[ 
\left(\frac{y\nu}{a}\right)^2-m^2\right]^{-s}\frac{\partial}{\partial y}\left[\log\left(y^{-\nu}\tilde{\Delta}^{TE}  ( y\nu)\right)\right]_{\nu\to\infty}\nonumber \\
& =:& C_0+C_1+C_2 
\eeq
with (in the following we perform our calculations up to $n_{max}=4$)
\beqs
\left[\log\left(y^{-\nu}\tilde{\Delta}^{TE}  ( y\nu)\right)\right]_{\nu\to\infty}  &\sim& \log\left\{y^{-\nu} \frac{1}{\sqrt{2\pi \nu}}{e^{\nu \eta}}{\left(1+y^2\right)^{1/4}}\left[-\left( 1+ \sum_{k=1}^{n_{max}} \nu ^{-k} v_k(t) \right)
+ \right.\right. \nonumber \\
& &\left.\left. {t}\left(  x_{\dd}-\frac{1}{2\nu}\right)\left(1+\sum _{k=1}^{n_{max}} \nu ^{-k} u_k(t) \right) \right]  \right\} \nonumber 
\eeqs
so that
\beqs
C_0&=& \frac{\sin(\pi s)}{\pi}\sum_{\nu=3/2}^\infty \nu \int_{m a/\nu}^\infty dy
\left[ \left(\frac{y\nu}{a}\right)^2-m^2\right]^{-s}\frac{\partial}{\partial y}
\log\left[y^{-\nu} \frac{1}{\sqrt{2\pi \nu }}{e^{\nu \eta}}{\left(1+y^2\right)^{1/4}}\right]  \nonumber \\
C_1 &=&\frac{\sin(\pi s)}{\pi}\sum_{\nu=3/2}^\infty \nu \int_{m a/\nu}^\infty dy \left[ 
\left(\frac{y\nu}{a}\right)^2-m^2\right]^{-s}\frac{\partial}{\partial y}\log[\frac{1}{x_{\dd}} ] \nonumber \\
C_2 & = &\frac{\sin(\pi s)}{\pi}\sum_{\nu=3/2}^\infty \nu \int_{m a/\nu}^\infty dy \left[ 
\left(\frac{y\nu}{a}\right)^2-m^2\right]^{-s} \left\{A_0(y)+
\frac{A_1(y) }{\nu}  +\frac{A_2(y)}{\nu^2}+\frac{A_3(y)}{\nu^3}+\frac{A_4(y)}{\nu^4}\right\},
\eeqs
where the functions $A_i(y)$, $(i=0,4)$ are given in the appendix A.

For the sake of clarity the calculations are developed in Appendix B and here we report the results only:
\beqs
C_0&=&   \frac{1}{4} \left(  a^{2 s} \zeta \left(2
   s-1,\frac{3}{2}\right)-\frac{  a^{2 s} \zeta \left(2
   s-2,\frac{3}{2}\right) \sin (\pi  s) \Gamma
   \left(s-\frac{1}{2}\right) \Gamma (-s)}{\pi ^{3/2}}\right)  \\
C_1 &=& \frac{11y_a}{48a} \\
C_2 &=& \frac{\sin(\pi s)}{\pi} a^{2 s} \left[\zeta \left(2 s-1,\frac{3}{2}\right)f_0^{TE}(s)+
\zeta \left(2 s,\frac{3}{2}\right)f_1^{TE}(s)+\zeta \left(2 s+1,\frac{3}{2}\right) f_2^{TE}(s)+ \right. \\
&&\left. \zeta \left(2 s+2,\frac{3}{2}\right)f_3^{TE}(s)+ \zeta \left(2 s+3,\frac{3}{2}\right)f_4^{TE}(s)\right] 
\eeqs
with the $f_i^{TE}(s)$ for $(i=0,4)$ given in appendix B.

expanding around $s=-1/2$  we finally obtain:
\beq
\zeta^{asym}_{TE}(-1/2) &=& \frac{1}{a}\Biggl\{0.107+0.229 y_a+0.127y_a^2 +
0.0706 \log(a)+\frac{0.0353}{s+1/2}+  \\
 & &y_a^2 \left(\frac{0.0624-0.0398 \log
   (a)}{s+1/2}-0.0398 \log ^2(a)+0.125 \log(a)-
   \frac{0.0199}{(s+1/2)^2}\right)\Biggr\}  \nonumber
\eeq

\subsection{TM-Modes}   

In the same manner we have
\beq
\zeta_{TM}^{asym}(s)
&=&\frac{\sin(\pi s)}{\pi}\sum_{\nu=3/2}^\infty \nu \int_{m a/\nu}^\infty dy\left[ 
\left(\frac{y\nu}{a}\right)^2-m^2\right]^{-s}\frac{\partial}{\partial y}\left[\log\left(y^{-\nu}\tilde{\Delta}^{TM}  ( y\nu)\right)\right]_{\nu\to\infty}\nonumber \\
& =:& D_0+D_1+D_2 
\eeq
with 
\beq
\left[\log\left(y^{-\nu}\tilde{\Delta}^{TM}  ( y\nu)\right)\right]_{\nu\to\infty}&=&
\log\left\{y^\nu \sqrt{\frac{ \nu}{2\pi}}{ e^{\nu \eta}}{\left(1+y^2\right)^{1/4}}
\left[ \sum _{k=1}^{\text{nmax}} \nu ^{-k} v_k(t)+1 +\right.\right. \nonumber \\
    & & \left.\left. \frac{t }{\nu }\left(\frac{1}{2}-
     \nu  \frac{y^2}{ x_{\dd}}\right)\left(\sum _{k=1}^{\text{nmax}} \nu ^{-k}
    u_k(t)+1\right)\right]\right\}  
\eeq
and
\beq
D_0 &=&  \frac{\sin(\pi s)}{\pi}\sum_{\nu=3/2}^\infty \nu \int_{m a/\nu}^\infty dy 
\left[ \left(\frac{y\nu}{a}\right)^2-m^2\right]^{-s}\frac{\partial}{\partial y}\log\left[y^\nu \sqrt{\frac{ \nu}{2\pi}}{ e^{\nu \eta}}{\left(1+y^2\right)^{1/4}}\right] = C_0 \nonumber  \\
D_1 &=& \frac{\sin(\pi s)}{\pi}\sum_{\nu=3/2}^\infty \nu \int_{m a/\nu}^\infty dy 
\left[ \left(\frac{y\nu}{a}\right)^2-m^2\right]^{-s}\frac{\partial}{\partial y}\log \left(1-\frac{y}{\sqrt{y^2+1}}\right)  \nonumber \\
D_2 & = &  \frac{\sin(\pi s)}{\pi}\sum_{\nu=3/2}^\infty \nu \int_{m a/\nu}^\infty dy 
\left[ \left(\frac{y\nu}{a}\right)^2-m^2\right]^{-s}\left\{ \frac{B_1(y) }{\nu}  +\frac{B_2(y)}{\nu^2}+\frac{B_3(y)}{\nu^3}+\frac{B_4(y)}{\nu^4} \right\} \nonumber 
\eeq
with the functions $B_i$, $(i=1,4)$  given in  the appendix A.
We find
\beqs
D_0 &=& C_0 \\
D_1&=&-\frac{a^{2 s} \zeta \left(2 s-1,\frac{3}{2}\right) \left(\sin (\pi  s)
   \Gamma \left(\frac{1}{2}-s\right) \Gamma (s)+\pi ^{3/2}\right)}{2 \pi
   ^{3/2}}  \\
D_2 &=& \frac{\sin(\pi s)}{\pi} a^{2 s} \left[  \zeta \left(2 s,\frac{3}{2}\right)f_1^{TM}(s)+
\zeta \left(2 s+1,\frac{3}{2}\right) f_2^{TM}(s)+ \zeta \left(2 s+2,\frac{3}{2}\right)f_3^{TE}(s)+ \right. \\
&&\left. \zeta \left(2 s+3,\frac{3}{2}\right)f_4^{TM}(s)\right]
\eeqs
with the$f_i^{TM}(s)$, $(i=1,4)$ defined in appendix B.
 Expanding aroud $s=-1/2$ we find:
\beqs
\zeta_{TM}^{asym}  &=& \frac{1}{a} \Bigl\{-0.170+0.266y_a^2-0.317y_a^4- 0.0554 \log ( a)-
\frac{0.0277}{s+1/2}+ \\
&&  y_a^2
   \left(\frac{0.130-0.119 \log ( a)}{s+1/2}-0.119 \log ^2( a)+
   0.259 \log( a)-\frac{0.0597}{(s+1/2)^2}\right)+ \\
& &    y_a^4 \left(0.130 \log ( a)+\frac{0.065}{s+1/2}\right)
   \Bigr\}
\eeqs
Thus the final result is (after having reintroduced the parameter $\mu$):
\beq
E_{Cas} &=& \frac{1}{a}\Bigl\{-0.063+0.229 y_a+0.393  y_a^2-
0.317y_a^4+   0.015 \log (\mu a)+\frac{0.008}{s+1/2}+ \nonumber \\
& &   y_a^2
   \left(\frac{0.192-0.159 \log (\mu a)}{s+1/2}-0.159 \log ^2(\mu a)+
   0.384 \log(\mu a)-\frac{0.0796}{(s+1/2)^2}\right)+\nonumber\\ 
& &y_a^4 \left(0.130 \log(\mu a)+\frac{0.065}{s+1/2}\right)  \Bigr\}\label{eq:ecasfin}
\eeq

\section{Discussion}

Formula (\ref{eq:ecasfin}) has many interesting features. First the structure of the divergencies: together with the usual terms depending on $\frac{1}{s+1/2}$ and $\log(\mu a)$ we find a new one: a second order pole in $(s+1/2)$ and the $\log^2(\mu a)$ term. This terms cannot be eliminated computing the Principal Part of the zeta functions as usually done  within the  zeta function regularization \cite{blau88}. What we expect, in general, is that these divergencies cancel when exterior field modes are included \cite{milton83,bord09}. However the peculiar (and interesting) feature of the SI approach is exactly the fact that one can perform all the calculation without any reference to exterior modes. 
So it would be desirable that surface terms renormalize the occurring divergencies but, on the other side, it cannot be excluded that  the renormalization procedure will induce further finite contributions to the calculated ones. In this respect the situation is similar to that of a shell of finite thickness in which one cannot give a satisfactory interpretation of the vacuum energy  \cite{bord09}, and further investigations are necessary.

More interesting, from the physical point of view, results the finite part of the energy:
\be
E_{Cas} = \frac{1}{a}\Bigl\{-0.063+0.229 y_a+0.393  y_a^2-
0.317y_a^4\Bigr\} \label{eq:ecasfin2}
\ee
Indeed it turns into negative values both for small, $y_a<0.2$, and large, $y_a>1.3$, values of $y_a$. In particular it is negative for $y_a=0$, but, it corresponds to a situation in which the SI of the material goes to one for every values of $\omega$, obviously an unrealistic one from the point of view of real materials.

Concerning the other range: $y_a>1.3$, it seems to indicate that for greater and greater values of $y_a$ the Casimir energy gets more and more negative.This is a very delicate point and it deserves deeper examination. 

It is true that, in doing the calculations,  we made no assumptions on the values of $ y_a$, but, as one can easily realize, the asymptotic expansion we did is non uniform with respect to the parameter $ y_a$ in the range $ y_a\in[0,\infty]$. This can be traced by observing that the two limits $ y_a\to\infty$ and $\nu\to\infty$ do not commute, so, once we did the expansion for $\nu\to\infty$ we are no more allowed to take $ y_a$ as large as we want.  Incidentally we note that this prevents us to take the limit $ y_a\to\infty$ in Eq. (\ref{eq:ecasfin}) that would give the Casimir energy for a perfect conducting sphere because $\lim_{ y_a\to\infty}{\cal{Z}}=0.$

To recover this limit, instead, we had to make the limit $ y_a\to\infty$ first and then to proceed with the asymptotic expansion with respect to $\nu$. Doing in this way we find 
\beq
E_{cas}|_{ y_a\to\infty} &=&\frac{1}{a}\Bigl[0.084+0.0081 \log(\mu a)+\frac{0.0040}{s+1/2} + \nonumber\\
& &\frac{1}{y_a}\Bigl(-0.194-0.0312 \log (\mu a)-\frac{0.0156}{s+1/2}\Bigr)\Bigr]
\eeq
in agreement  with \cite{lese96,milton83}.
We note that even in this case, the correction due to $ y_a$ allows for a negative Casimir energy for $ y_a<2.3$ (but, perhaps, this must be considered a too small value with respect to the assumption $y_a\to\infty$). 

In any case, even though it would be very difficult, maybe impossible, for real materials to fulfill the conditions to have negative Casimir energy,  these results can be very interesting from the point of view of the application of the MIT bag model to the confinement of quarks into the hadrons. Indeed, to recover the zero-point contribution of the (static) MIT bag energy, $E(a)\simeq-\frac{1.84}{a}$, obtained from the phenomenological bag model fits for the spectrum of hadronic particles \cite{plunien}, we need simply $y_a\simeq1.823$ (where we used Eq. (\ref{eq:ecasfin2}) and considered the fact that the gluons are eight).
Our result  {\em simply } tells us that probably the hadron surface is not an ideal conductor with respect to the color. 

However a fully satisfactory application to QCD has to face with stronger problems as for example the self interaction of the gluons and the non abelian nature of the gauge group with the consequent change of  the propagator of the theory 
\cite{dudal08,pattan10} that could eventually modify the result. In this respect this is only a first step.

It would be very interesting, in our opinion,
to find a uniform asymptotic expansion for $ y_a\in[0,\infty]$ so to have the possibility of computing the two limits $ y_a\to0$ and $ y_a\to\infty$ on the same formula and to extend the range of validity of (\ref{eq:ecasfin2})  to  $y_a\to \infty$ too.  Also interesting, from the physical point of view, would be  to extend the same approach to more general Surface Impedance functional forms.

\section*{Acknowledgments}
The author is grateful to Giampiero Esposito for comments and suggestions.

\section*{Appendix A}
\beqs
A_0(y) &=& \frac{1}{\sqrt{y^2+1}}-\frac{y}{y^2+1}\\
A_1(y) &=&\frac{y \left\{y \left[y \left(8 y
   \left(\sqrt{y^2+1}-y\right)-21\right)+20
   \sqrt{y^2+1}\right]-8\right\}+4 \sqrt{y^2+1}}{8
   \left(y^2+1\right)^{5/2}}\\
A_2(y)  &=& \frac{16 y^8+56 y^6+73 y^4+10 y^2+2}{8 \left(y^2+1\right)^{7/2}}-
   \frac{y \left(16 y^8+64 y^6+97y^4+38 y^2+4\right)}{8 \left(y^2+1\right)^4}
   -\frac{\delta_\nu^2}{2 y^2 \sqrt{y^2+1}}\\
A_3(y)  &=&\frac{1}{128 \left(y^2+1\right)^{11/2}}\Bigl[-80 y-512y^{13}
-2720 y^{11}-5808 y^9-6193 y^7-1520 y^5-112 y^3+ \\
   & &\sqrt{y^2+1}\left(512 y^{12}+2464 y^{10}+4640 y^8+4176 y^6-160 y^4+288
   y^2+16\right)\Bigr]+\\
   & &\frac{\delta_\nu^2 }{4 y^2 \left(y^2+1\right)^{5/2}}\left(-4 y^7-10 y^5-6 y^3+
   \sqrt{y^2+1} \left(4 y^6+8 y^4+y^2-1\right)\right) \\
A_4(y)  &=& \frac{3 \delta_\nu^4}{8 y^4 \sqrt{y^2+1}}+\frac{\delta_\nu^2
   \left[\left(64 y^8+224 y^6+272 y^4+91 y^2+2\right) y^2\right]}{16 y^2
   \left(y^2+1\right)^{7/2}}+ \\
& &\frac{\delta_\nu^2\left[-8 \sqrt{y^2+1}
   \left(8 y^6+24 y^4+23 y^2+3\right) y^3-2\right]}{16 y^2
   \left(y^2+1\right)^{7/2}}+ \frac{  (1024 y^{16}+7296 y^{14}+22080
   y^{12})y^2}{128\left(y^2+1\right)^{15/2}}\\
& &\frac{  (36336 y^{10}+33763 y^8+8007 y^6-1512 y^4+ 
    3780 y^2+72)y^2}{128\left(y^2+1\right)^{15/2}} -\frac{ 8}{128\left(y^2+1\right)^{15/2}}- \\
  & &\frac{ 4 \left(256 y^{16}+1696 y^{14}+4704 y^{12}+6928
   y^{10}+5461 y^8-42 y^6+208 y^4+356 y^2+4\right) y}{128\left(y^2+1\right)^{7}}\\
 \eeqs

\beqs
B_1(y) &=& \frac{1-y^2}{2 \left(y^2+1\right)^2}-\frac{5 y^3}{8
   \left(y^2+1\right)^{5/2}} \\
B_2(y)   &=& \frac{y \left(-9 y^5+5 y^3+\sqrt{y^2+1} \left(-9 y^4+10 y^2+4\right)+12
   y\right)-2}{8 \left(y^2+1\right)^{9/2}}-\frac{{ \delta_\nu}^2}{2 y^2
   \sqrt{y^2+1}} \\
B_{3}(y)  &=& \frac{1}{4} { \delta_\nu}^2 \left(\frac{2
   y}{\left(y^2+1\right)^{3/2}}+\frac{2
   y^4+y^2+1}{\left(y^3+y\right)^2}\right)+\frac{-25 y^6+70 y^4-24
   y^2+1}{8 \left(y^2+1\right)^5}+ \\
 & & \frac{-401 y^7+928 y^5+112 y^3-112y}{128 \left(y^2+1\right)^{11/2}}\\
B_4(y)  &=& { \delta_\nu}^4 \left(\frac{1}{y^3}+\frac{8 y^2+9}{8 y^4
   \sqrt{y^2+1}}\right)+{ \delta_\nu}^2 \left(\frac{y \left(y^2-3\right)}{2
   \left(y^2+1\right)^3}+\frac{8 y^6-21 y^4-6 y^2-2}{16 y^2
   \left(y^2+1\right)^{7/2}}\right)+  \\
 & &  \frac{5980 y^6-1363 y^8-4292 y^4+512
   y^2-8}{128 \left(y^2+1\right)^{13/2}}+\frac{1330 y^7-341 y^9-376
   y^5-316 y^3+36 y}{32 \left(y^2+1\right)^7}
\eeqs
   
\section*{Appendix B}   
In the following all the relevant integrations can be obtained as limiting cases of the following formula 
\beqs
 \int_{m a/\nu}^\infty dy\left[ 
\left(\frac{y\nu}{a}\right)^2-m^2\right]^{-s} \frac{y^b}{(1+y^2)^c} &=&\frac{1}{2 m^{2s}}
\left( \frac{a m}{\nu}\right)^{b-2 c+1} 
 B\left(1-s,c+s-\frac{1+b}{2}\right)  \\
 & &    _2F_1\left(c,c+s-\frac{1+b}{2},+c+\frac{1-b}{2},-\frac{\nu ^2}{a^2 m^2}\right); \\
\eeqs
with $B(x,y)=\frac{\Gamma(x)\Gamma(y)}{\Gamma(x+y)}$ the beta function and $_2F_1(a,b,c;x)$ the hypergeometric function \cite{abste}. Indeed one easly realizes that quite often it is possible to put $m=0$ from the very beginning and in this case one can use directly
\beqs
 \int_{0}^\infty dy
\left(\frac{y\nu}{a}\right)^{-2s} \frac{y^b}{(1+y^2)^c} =\frac{\left(\frac{\nu ^2}{a^2}\right)^{-s} 
\Gamma \left(\frac{ b-2 s+1}{2}\right) 
\Gamma \left(c+s-\frac{1+b}{2}\right)}{2 \Gamma(c)}
\eeqs
In this way we find for the TE-modes:
\beq
C_0 &=& \sum_{\nu=3/2}^\infty\frac{ \nu   \sin (\pi  s)}{\pi }\int_{0}^\infty
\left(  \frac{\nu ( \sqrt{1+y^2} -1)}{y}+\frac{  y}{2
   \left(y^2+1\right)} \right) 
   \left(\frac{y\nu}{a}  \right)^{-2s} dy \nonumber \\
 &=&  \sum_{\nu=3/2}^\infty\frac{ \nu   \sin (\pi  s)}{\pi }
   \frac{\left(\frac{\nu }{a}\right)^{-2 s} \left(\pi ^{3/2}  
   \csc (\pi  s)-\nu    \Gamma \left(s-\frac{1}{2}\right)
    \Gamma (-s)\right)}{4 \sqrt{\pi }} \nonumber \\
&=&   \frac{1}{4} \left(  a^{2 s} \zeta \left(2
   s-1,\frac{3}{2}\right)-\frac{  a^{2 s} \zeta \left(2
   s-2,\frac{3}{2}\right) \sin (\pi  s) \Gamma
   \left(s-\frac{1}{2}\right) \Gamma (-s)}{\pi ^{3/2}}\right).
 \eeq
In treating $C_1$ we have to introduce a spurious exponent $\alpha$ for the $y$ so to avoid the divergencies when summing over $\nu$, thus we define
\beq
C_1 &=& \lim_{ \alpha\to1} \sum_{\nu=3/2}^\infty\frac{ \nu   \sin (\pi  s)}{\pi }\int_{0}^\infty
\frac{-\nu ^2 y^{{ \alpha}}}{\nu ^2 y^2+{ \delta_\nu}^2}
   \left(\frac{y\nu}{a}\right)^{-2s}dy   \\
& = & \lim_{ \alpha\to1} \sum_{\nu=3/2}^\infty  -
 \frac{1}{2} a^{2 s} \nu ^{2-{ \alpha}} \sin (\pi  s)
  y_a^{{ \alpha}-2 s-1} \sec \left(\frac{1}{2} \pi 
   ({ \alpha}-2 s)\right)  \nonumber \\
& = & \lim_{ \alpha\to1} -\frac{1}{2} a^{2 s} \zeta \left({ \alpha}-2,\frac{3}{2}\right)
   \sin (\pi  s) y_a^{{ \alpha}-2 s-1} \sec \left(\frac{1}{2} \pi  ({ \alpha}-2 s)\right)  \nonumber
\eeq
And  for $C_2$:
\beqs
C_2 & = &\frac{\sin(\pi s)}{\pi}\sum_{\nu=3/2}^\infty \nu \int_{m a/\nu}^\infty dy \left[ 
\left(\frac{y\nu}{a}\right)^2-m^2\right]^{-s} \left\{ A_0(y)+\frac{A_1(y) }{\nu}  +\frac{A_2(y)}{\nu^2}+\frac{A_3(y)}{\nu^3}+\frac{A_4(y)}{\nu^4} \right\} \\
&=&\frac{\sin(\pi s)}{\pi} a^{2 s} \sum_{\nu=3/2}^\infty \nu ^{-2 s+1}\left[f_0^{TE}(s)+\frac{f_1^{TE}(s)}{\nu}+ \frac{f_2^{TE}(s)}{\nu^2}+ \frac{f_3^{TE}(s)}{\nu^3}+ 
\frac{f_4^{TE}(s)}{\nu^4}\right] \\
& =& \frac{\sin(\pi s)}{\pi} a^{2 s} \left[\zeta \left(2 s-1,\frac{3}{2}\right)f_0^{TE}(s)+
\zeta \left(2 s,\frac{3}{2}\right)f_1^{TE}(s)+
\zeta \left(2 s+1,\frac{3}{2}\right) f_2^{TE}(s)+  \right. \\
&&\left. \zeta \left(2 s+2,\frac{3}{2}\right)f_3^{TE}(s)+\zeta \left(2 s+3,\frac{3}{2}\right)f_4^{TE}(s)\right] 
\eeqs

with
\beqs
f_0^{TE}(s)&=&\frac{1}{2} a^{2 s} \left(\frac{\Gamma \left(\frac{1}{2}-s\right) \Gamma
   (s)}{\sqrt{\pi }}-\frac{\pi}{  \sin (\pi  s)}\right)\\
f_1^{TE}(s) &=&-\frac{a^{2 s} \zeta \left(2 s,\frac{3}{2}\right) \left(12 \pi ^{3/2} s
   \sec (\pi  s)+(5 (3-2 s) s+7) \Gamma (1-s) \Gamma
   \left(s-\frac{1}{2}\right)\right)}{24 \sqrt{\pi }}\\
f_2^{TE}(s)   &=& s a^{2 s} \Bigl[\frac{\pi  ((13-5 s) s+16) }{32\sin (\pi s)} -\\
& &\frac{\left(12 y_a^2 (s-1)+40 s^4-28 s^3-58 s^2+s+9\right)
   \Gamma \left(-s-\frac{1}{2}\right) \Gamma (s-1)}{48 \sqrt{\pi  }}\Bigr]\\
f_3^{TE}(s)  &=& a^{2 s} \Bigl\{- \frac{1}{384 \cos (\pi s)} \pi  s \left[96 y_a^2+(2 s+1) (4 s (15 s-7)-169)\right]+ \\
& &  \frac{1}{60480 \sqrt{\pi }}\Bigl[(2 s+1) \bigl(-7560 y_a^2 (2 s-3)+s (s (s (4 s
   (1105 s-7184)+18431)+\\
& &  69542)- 62271)+1764\bigr) \Gamma (1-s) \Gamma
   \left(s-{3/2}\right)\Bigr]\Bigr\} \\
 f_4^{TE}(s) &=&  a^{2 s} \Bigl\{\frac{\pi  \left(768 y_a^2 (2 s+3)+s (s (s ((621-113 s)
   s+1691)-2253)-4362)-384\right) }{3072 \sin (\pi  s)}+  \\
 & &  \frac{y_a^2
   (s (s (77-20 (s-1) s)+4)-9)  \Gamma \left(-s-\frac{1}{2}\right)
   \Gamma (s-1)}{48 \sqrt{\pi }}+ \\
& &   \frac{1}{60480 \sqrt{\pi }} [\Bigl(s [s \left(s \left(8 s
   \left(s \left(2210 s^2-5968
   s-10351\right)+37085\right)+3385\right)-379346\right)+\\
 & & 129783]+ 17010\Bigr)  \Gamma \left({1/2}-s\right) \Gamma(s-2)]\Bigr\}\\
\eeqs

For the TM-modes we get:

\beqs
D_0 &=&C_0 \\
D_1 &=& \frac{\sin(\pi s)}{\pi}\sum_{\nu=3/2}^\infty \nu \int_{0}^\infty dy 
\left(\frac{y\nu}{a}\right)^{-2s}\left[-\frac{y}{y^2+1}-\frac{1}{\sqrt{y^2+1}}\right]  \nonumber \\
 &=& \frac{\sin(\pi s)}{\pi}\sum_{\nu=3/2}^\infty\frac{a^{2 s} 
 \nu^{1 - 2 s}}{2} \left(-\frac{\pi}{  \sin (\pi  s) }-\frac{\Gamma
   \left(\frac{1}{2}-s\right) \Gamma (s)}{\sqrt{\pi }}\right)\\
&=&-\frac{a^{2 s} \zeta \left(2 s-1,\frac{3}{2}\right) \left(\sin (\pi  s)
   \Gamma \left(\frac{1}{2}-s\right) \Gamma (s)+\pi ^{3/2}\right)}{2 \pi
   ^{3/2}}  \\
  D_2 & = &  \frac{\sin(\pi s)}{\pi}\sum_{\nu=3/2}^\infty \nu \int_{m a/\nu}^\infty dy 
\left[ \left(\frac{y\nu}{a}\right)^2-m^2\right]^{-s}\left\{ \frac{B_1(y) }{\nu}  +\frac{B_2(y)}{\nu^2}+\frac{B_3(y)}{\nu^3}+\frac{B_4(y)}{\nu^4} \right\}  \\
 &=& \frac{\sin(\pi s)}{\pi} a^{2 s}\sum_{\nu=3/2}^\infty \nu ^{-2 s+1}\left[\frac{1}{\nu}f_1^{TM}(s)+ \frac{1}{\nu^2}f_2^{TM}(s)+ \frac{1}{\nu^3}f_3^{TM}(s)+\frac{1}{\nu^4}f_4^{TM}(s)\right]  \\
& =& \frac{\sin(\pi s)}{\pi} a^{2 s} \left[
\zeta \left(2 s,\frac{3}{2}\right)f_1^{TM}(s)+\zeta \left(2 s+1,\frac{3}{2}\right) f_2^{TM}(s)+ \zeta \left(2 s+2,\frac{3}{2}\right)f_3^{TM}(s)+ \right. \\
&&\left. \zeta \left(2 s+3,\frac{3}{2}\right)f_4^{TM}(s)\right] 
\eeqs 

with
\beqs
f_1^{TM}(s) &=&\frac{1}{12 \sqrt{\pi }}\left[\frac{6 \pi ^{3/2} s }{\cos (\pi  s)}-5 \Gamma (2-s) \Gamma\left(s+{1/2}\right)\right] \\
f_2^{TM}(s) &=&\frac{1}{96 \sqrt{\pi }}\Bigl[2 \left(3-12 { y_a}^2+2 s \left(20 s^2+6 s+1\right)\right)
   \Gamma \left(-s-{1/2}\right) \Gamma (s+1)+ \\
& &   3 \pi ^{3/2} (13-5 s)s^2 \frac{1}{\sin (\pi  s)}  \Bigr] \\
f_3^{TM}(s) &=&\frac{1}{120960 \sqrt{\pi }}
\Bigl[315 \pi ^{3/2} s \left(23-96 { y_a}^2+2 s \left(60 s^2+2
   s+9\right)\right)\frac{1}{ \cos (\pi  s)}+ \\
 & &  8 \left(7560 { y_a}^2+(2 s+1)
   (s (131+s (1105 s-4974))-672)\right) \Gamma (1-s) \Gamma
   \left(s+{1/2}\right)\Bigr] \\
f_4^{TM}(s) &=& -\frac{\sin(\pi s) a^{2 s} \zeta \left(2 s+3,\frac{3}{2}\right)}{967680 {\pi }^{3/2}} \Bigl\{
\frac{315 \pi ^{3/2} s}{\sin(\pi s)} \left[768 { y_a}^2 (2 s+1)+   \right. \\
& & \left.  s (s+1) (s (s (113 s-734)+3)-246)\right] - \\
& & 4 \Gamma \left(-s-\frac{3}{2}\right) \Gamma (s+1)\Bigl[15120 { y_a}^4 (s-3)+
2520 { y_a}^2 (2 s+3) \left(20 s^3-5 s+3\right)- \\
& &  (s+1) (2 s+1) (2 s+3) (8 s (s   (2 s (1105 s-774)+1363)-177)+945)\Bigr]
\Bigr\}
\eeqs

\end{document}